\documentclass[runningheads]{llncs}
\usepackage{epsfig}
\usepackage{graphicx}
\usepackage{subfigure}
\usepackage{amsmath}
\usepackage{amssymb}

\usepackage{amsthm}

\usepackage{amsfonts}
\usepackage{mathrsfs}
\usepackage{booktabs}
\usepackage{multirow, eucal}
\usepackage{color}
\usepackage{cite}
\usepackage{wrapfig,lipsum,booktabs}
\usepackage{caption}
\usepackage[vlined,ruled,linesnumbered]{algorithm2e}

\def\bI{{\bf I}}

\def\bR{{\bf R}}
\def\bS{{\bf S}}
\def\bT{{\bf T}}

\def\mL{{\mathcal L}}

\begin{document}
\title{Retinal Image Segmentation with a Structure-Texture Demixing Network \thanks{M. Tan (mingkuitan@scut.edu.cn) and Y. Xu (ywxu@ieee.org) are the corresponding authors.}}

\author{
Shihao Zhang$^{1,2*}$ \and
Huazhu Fu$^{3}$ \and
Yanwu Xu$^{4\dagger}$ \and
Yanxia Liu $^{1}$\and
Mingkui Tan$^{1\dagger}$
}

\authorrunning{S. Zhang et al.}

\institute{South China University of Technology, Guangzhou, China \and
Pazhou Lab, Guangzhou, China \and
Inception Institute of Artificial Intelligence, Abu Dhabi, UAE  \and
Cixi Institute of Biomedical Engineering, Ningbo Institute of Materials Technology and Engineering, Chinese Academy of Sciences, Ningbo, China
%Baidu, inc., China
}

\maketitle
\thispagestyle{empty}

\begin{abstract}
Retinal image segmentation plays an important role in automatic disease diagnosis. This task is very challenging because the complex structure and texture information are mixed in a retinal image, and distinguishing the information is difficult.
Existing methods handle texture and structure jointly, which may lead biased models toward recognizing textures and thus results in inferior segmentation performance. To address it, we propose a segmentation strategy that seeks to separate structure and texture components and significantly improve the performance. To this end, we design a structure-texture demixing network (STD-Net) that can process structures and textures differently and better.
Extensive experiments on two retinal image segmentation tasks (\textit{i.e.}, blood vessel segmentation, optic disc and cup segmentation) demonstrate the effectiveness of the proposed method.


\end{abstract}

\section{Introduction} \label{Sec:Introduction}
Retinal image segmentation is important in automatic disease diagnosis \cite{jelinek2009automated,hancox1999optic}.
For example, retinal vessels are correlated to the severity of diabetic retinopathy, which is a cause of blindness globally \cite{jelinek2009automated}.
Moreover, the optic disc (OD) and optic cup (OC) are used to calculate the cup-to-disc-ratio (CDR), which is the main indicator for glaucoma diagnosis \cite{hancox1999optic}.
However, retinal image segmentation is often extremely challenging because retinal images often contain complex texture and structure information, which is different from general natural images.

Recently, deep neural networks (DNNs) have shown a strong ability in image segmentation with remarkable improvements \cite{ronneberger2015u,Gu2019CE,zhang2019ETNet,zhang2019whole,zhang2020collaborative}.
However, existing methods are strongly biased toward recognizing textures rather than structures \cite{geirhos2018imagenet} since they handle the two types of information jointly.
As a result, tiny structures that are very similar to textures will be misclassified.
Therefore, separately processing the structure and texture information in a retinal image is necessary.
Structure-texture demixing is an essential operation in image processing that has been extensively utilized in many computer vision tasks, including image enhancement \cite{guo2016lime}, optical flow \cite{revaud2015epicflow} and image stylization \cite{gastal2011domain}.
However, the application of a structure-texture demixing operation in retinal image segmentation remains an open question.

Existing structure-texture demixing methods cannot adequately distinguish the boundary
structures from textures, because they may have similar statistical properties\cite{xu2012structure, kim2018structure}.
The texture component will inevitably contain structure information.
Therefore, the structure information is not
fully exploited by these methods, which produces inferior segmentation results.

In this paper, we propose a \textbf{Structure-Texture Demixing Network (STD-Net)} that decomposes the image into a structure component and a texture component.
Note that,
the structure and texture components have different properties and need to be treated differently. We exploit two types of networks to treat them differently.
The structure component mainly contains smooth structures, while the texture component mainly contains high-frequency information. Thus the structure component is suitable for processing by representative networks, and the texture component is easily overfitted, a shallower network is a better choice.
We conduct extensive experiments for two tasks: vessel segmentation using the DRIVE dataset, and optic disc and cup segmentation using the ORIGA and REFUGE datasets. The results demonstrate the effectiveness of our method.

The contributions of this paper are listed as follows:
1) We propose a segmentation strategy that demix a retinal image into structure and texture components. This strategy can be applied to any segmentation framework to improve its performance.
2) We design a structure-texture demixing network (STD-Net) that can process structures and textures differently and better.
3) Extensive experiments for two retinal image segmentation tasks demonstrate the effectiveness of the proposed strategy.

\begin{figure*}[!t]
	\centering
	\includegraphics[width=0.9\linewidth]{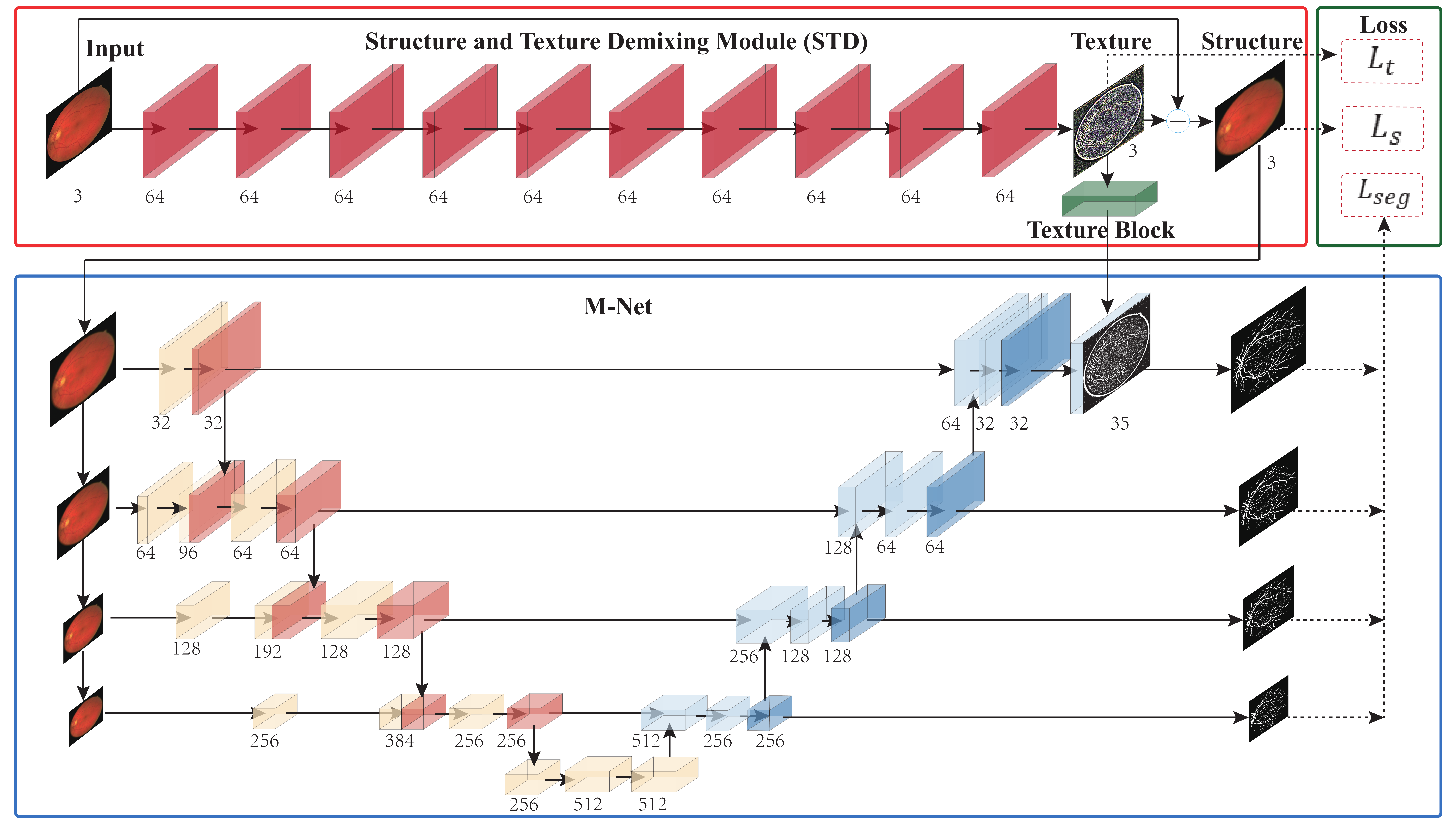}
	\caption{Overview of the proposed STD-Net. Built on the M-Net \cite{fu2018joint} as a backbone, STD-Net decomposes the input image into structure and texture components. The structure component serves as the input of M-Net to recover the boundary structures using the texture information extracted by a \textit{texture block} (refer to Fig. \ref{figBoundaryBlock}). The operator $\small{\textcircled{-}}$ represents the minus operation. The functions $\mL_t$, $\mL_s$ and $\mL_{seg}$ represent the texture loss,  structure loss, and  segmentation loss, respectively.
}
	\label{figall}
\end{figure*}

\section{Methodology} \label{Sec:Methodology}
We illustrate the overview of our proposed STD-Net in Fig. \ref{figall}. STD-Net decomposes an input image to a structure component and texture component. The structure component corresponding to the main object (smoothed part), and the texture component contains fine-grained details (almost periodic textures, noise).
The segmented object's primary information is contained in the structure component. We choose M-Net~\cite{fu2018joint} to process the structure component. The segmented object's detailed information is contained in the texture component, such as the boundaries. We propose a texture block to process the texture component. Details are provided in the following section.

\subsection{Structure-Texture Demixing Loss Function}
\label{SbSec:M1}

The structure-texture demixing module decomposes an image into a structure component and texture component by two types of loss functions, namely the structure loss and the texture loss.
The structure and the texture loss demix images by penalizing structures and textures differently. The different penalizes are based on statistical priors that structures and textures receive different penalty under some loss functions.

Given the input image $\bI$, the structure-texture demixing (STD) module aims to decompose $\bI$ into two components: $\bI\rightarrow \bS + \bT$, where $\bS$ and $\bT$ represent the structure component and texture component, respectively. This decomposition can be formulated as the following optimization problem:
\begin{equation}
\min \limits_{\bS,\bT} \lambda \mL_{s}(\bS) + \mL_{t}(\bT),
\label{equ1}
\end{equation}
where $\bS=\bI-\bT$, $\mL_s$ is the structure loss function and $\mL_t$ is the texture loss function, which leads $\bS$ and $\bT$ to different statistical properties, that is, for the structure component $\mL_s(\bS) \ll \mL_t(\bS)$ and for the texture component $\mL_s(\bT) \gg \mL_t(\bT)$. The constant $\lambda$ is the balancing parameter.

The total variation (TV) \cite{rudin1992nonlinear} is one of the most popular structure priors; we exploit it as the structure loss function $\mL_s$:

\begin{equation}
\mL_s(\bS) = \sum_{i,j}||(\nabla \bS)_{i,j}||_2,
\label{equ_TV}
\end{equation}
where $\nabla$ is the spatial gradient operator. Using the TV, various demixing methods have been proposed, e.g., TV-$\triangle^{-1}$ \cite{aujol2006structure}, TV-L2 \cite{rudin1992nonlinear}, and TV-L1 \cite{alliney1997property}. The L1-norm is more suitable for structure-texture demixing\cite{aujol2006structure}. Specifically, the texture loss function can be defined as follows:
\begin{equation}
\mL_t(\bT) = ||\bT||_1.
\label{equ_TV}
\end{equation}

We employ the cross-entropy function $\mL_{seg}$ as the segmentation loss function.
The final loss function $\mL_{total}$ is defined as:
\begin{equation}
\begin{aligned}
\mL_{total} (\bI, \bT, \bR) & = \mL_{seg}(\bR) + \mu(\mL_t(\bT) + \lambda \mL_{s}(\bS))\\
 &= \mL_{seg}(\bR) + \mu(\mL_t(\bT) + \lambda \mL_{s}(\bI-\bT)),
\end{aligned}
\label{equ_TV}
\end{equation}
where $\mu$ and $\lambda$ are trade-off parameters, and $\bR$ is the segmentation result.

\setlength{\abovecaptionskip}{0.cm}
\setlength{\belowcaptionskip}{-0.cm}
\begin{figure*}[!t]
	\centering
	\includegraphics[width=0.9\linewidth]{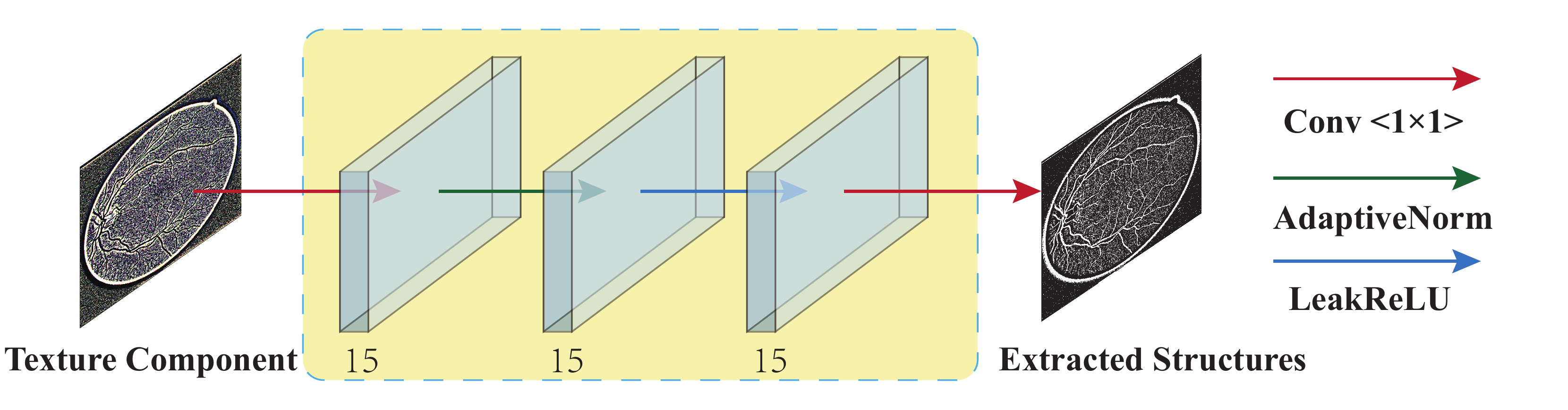}
	\caption{ Architecture of the texture block. The texture block is utilized to recover the falsely demixed structures and reduce the texture influence. }
	\label{figBoundaryBlock}
\end{figure*}

\subsection{Structure-Texture Demixing Module}
\label{SbSec:M2}
We show the architecture of the proposed Structure-Texture Demixing (STD) Module in Fig. \ref{figall}. First, we apply STD to extract the texture component. Second, we obtain the structure component by subtracting the texture component from the input image. In this way, we confirm that $\bI = \bS + \bT$. The STD consists of 10 convolutional layers with Leak ReLU to extract texture features. The extracted texture features are also serves as the input of the texture block.

\textbf{Texture Block:} The texture block is a component of STD. Because some structures, especially the boundary structures, may receive similar penalties from the structure loss and texture loss,
they may be misclassified as the texture components.
While these structures in texture component are important for segmentation, the textures and noises will affect the segmentation performance. To address it, the texture block is designed to extract boundaries and reduce the influence of textures and noises. Considering the limited amount of information in the texture component and a deep model may overfit, we design a very shallow network as the texture block.
Fig. \ref{figBoundaryBlock} shows the architecture of the texture block, which contains two convolution layers, an adaptive normalization layer~\cite{ogasawara2010adaptive} and a leaky ReLU layer.

\setlength{\abovecaptionskip}{0.cm}
\setlength{\belowcaptionskip}{-0.cm}
\begin{figure}[!t]
\centering
\subfigure[Input image] {
 \label{fig:a}
\includegraphics[width=0.2\columnwidth]{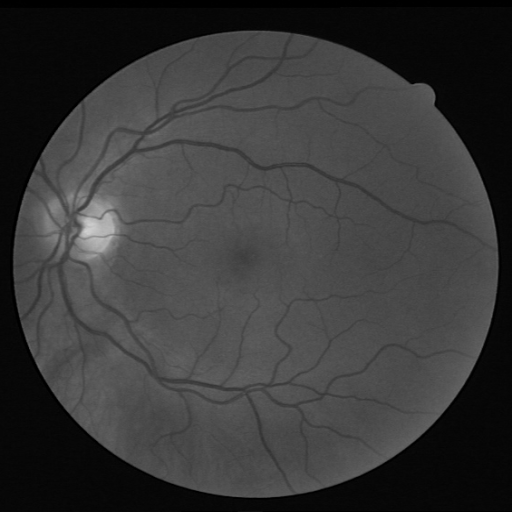}
}
\subfigure[Structure map] {
\label{fig:b}
\includegraphics[width=0.2\columnwidth]{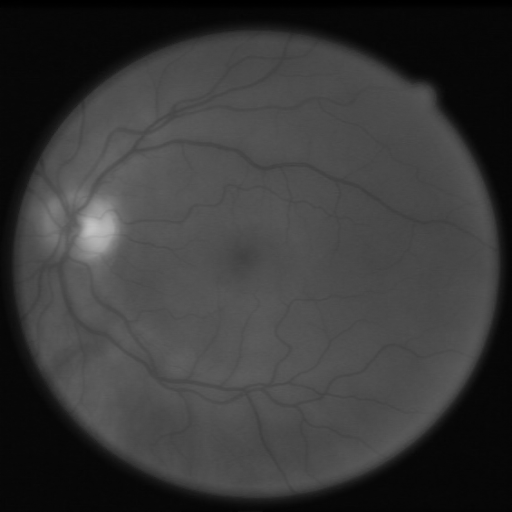}
}
\subfigure[Texture map] {
\label{fig:c}
\includegraphics[width=0.2\columnwidth]{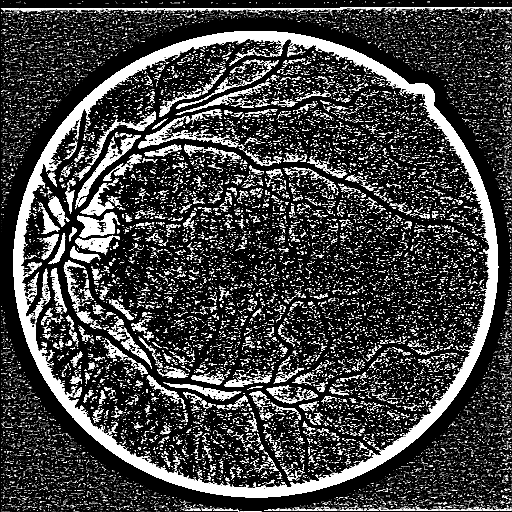}
}
\subfigure[E-structures] {
\label{fig:c}
\includegraphics[width=0.2\columnwidth]{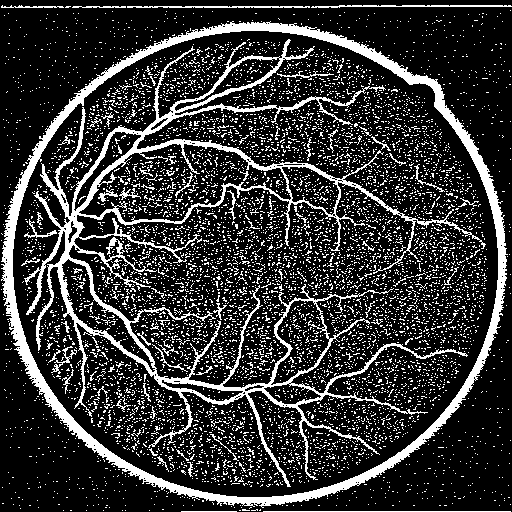}
}
\caption{Visualization of components in STD.
Comparing (a) and (b), using texture loss, the demixed structure component maintains most of the smooth structure information and filters out many high-frequency texture noises.
As shown in (c), the texture component mainly contains high-frequency information, which is a mixture of textures and boundary structures.
Comparing (c) and (d), the texture block clearly helps to extract structures in the texture component, while filtering out high-frequency textures.
}
\label{figExtractCompare}
\end{figure}

Fig. \ref{figExtractCompare} shows the visualization of the demixed structure,  demixed texture, and E-structures extracted by the texture block. To help observe more clearly, we only display the green (G) channels (RGB image).
The extracted structure component mainly contains smooth structures and the texture component mainly contains high-frequency information.
With the proposed texture block, we strengthen the structure information in the texture component and reduce the high-frequency textures.

\section{Experiments}
\label{Sec:Experiments}
In this paper, we evaluate our method in vessel segmentation and optic disc/cup segmentation from retina fundus images.
We train our STD-Net using Adam with a learning rate of 0.001. The batch size is set to 2. The balancing parameters $\lambda$ and $\mu$ are set to 1 and 0.001 respectively.

\subsection{Vessel Segmentation on DRIVE}
\label{Sec:DRIVEResult}

We conduct vessel segmentation experiments with DRIVE to evaluate the performance of our proposed STD-Net. The Digital Retinal Images for Vessel
Extraction (DRIVE) dataset \cite{staal2004ridge}  contains 40 colored fundus
images (20 training images and 20 testing images),
which are obtained from a diabetic retinopathy screening program in the Netherlands.
We resize the original images to $512 \times 512$ as inputs.
Following the previous work \cite{zhang2019attention}, we employ Specificity (Spe), Sensitivity (Sen), Accuracy (Acc), intersection-over-union(IOU), and Area Under ROC (AUC) as measurements.

\begin{table}[!t]
	\small
	\centering
	\caption{Quantitative comparison of segmentation results with DRIVE}
	    \begin{tabular}{cccccc}
		\toprule
		Method & Acc~~ & AUC~~ & Sen~~ & Spe~~ & IOU~~ \\
		\midrule
        Li \cite{li2016cross} & 0.9527~~ & 0.9738~~ & 0.7569~~ & 0.9816~~ & $-~~$ \\
        Liskowski \cite{liskowski2016segmenting} & 0.9535~~ & 0.9790~~ & 0.7811~~ & 0.9807~~& $-~~$ \\
		MS-NFN \cite{wu2018multiscale} & $0.9567~~$ & $0.9807~~$ & $0.7844~~$ & $0.9819~~$ & $-~~$ \\
        AG-Net \cite{zhang2019attention}  & $0.9692~~$ & $0.9856~~$ & $0.8100~~$   &  $0.9848~~$ & $0.6965~~$ \\
		U-Net \cite{ronneberger2015u}  & $0.9681~~$ & $0.9836~~$ & $0.7897~~$ & $0.9854~~$ & $0.6834~~$ \\
		M-Net \cite{fu2018joint} & $0.9674~~$ & $0.9829~~$ & $0.7680~~$ & $\mathbf{0.9868}~~$ & $0.6726~~$ \\
        STD-Net  & $\mathbf{0.9695}~~$ & $\mathbf{0.9863}~~$ & $\mathbf{0.8151}~~$ & $0.9846~~$ & $\mathbf{0.6995}~~$ \\
		\bottomrule
	\end{tabular}
	\label{DRIVE}
\end{table}

We compare our STD-Net with several state-of-the-art methods, including Li~\cite{li2016cross}, Liskowski~\cite{liskowski2016segmenting},  MS-NFN~\cite{wu2018multiscale},U-Net~\cite{ronneberger2015u}, M-Net~\cite{fu2018joint}, and AG-Net~\cite{zhang2019attention}.
Li~\cite{li2016cross} redefines the segmentation task as cross-modality data transformation from a retinal image to a vessel map, and outputs the label map of all pixels instead of a single label of the center pixel. Liskowski~\cite{liskowski2016segmenting} trains a deep neural network with samples that were preprocessed with global contrast normalization and zero-phase whitening and augmented using geometric transformations and gamma corrections.
MS-NFN~\cite{wu2018multiscale} generates multi-scale feature maps with an `up-pool' submodel and a `pool-up' submodel.
U-Net~\cite{ronneberger2015u} applies a contracting path to capture context and a symmetric expanding path to enable precise localization.
M-Net~\cite{fu2018joint} introduces multi-input and multi-output to learn hierarchical representations.
AG-Net~\cite{zhang2019attention} proposes a structure sensitive expanding path and incorporates it into M-Net.

Table~\ref{DRIVE} shows the performances of different methods for DRIVE. Based on the results, for the four metrics AUC, Acc, Sen, and IOU, the proposed STD-Net achieves the highest value. STD-Net outperforms the backbone M-Net by 0.0021, 0.0034, 0.0471 and 0.0269 in terms of Acc, AUC, Sen, and IOU, respectively. Note that the proposed STD-Net achieves a much higher Sen score than M-Net, which shows that our structure-texture demixing mechanism improves the structure detection ability of models.
\setlength{\abovecaptionskip}{0.cm}
\setlength{\belowcaptionskip}{-0.cm}
\begin{figure}[!t]
\centering
\subfigure[Input] {
 \label{fig:a}
\includegraphics[width=0.118\columnwidth]{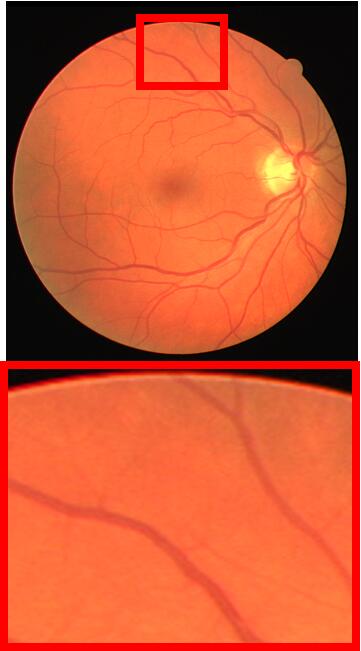}
}
\subfigure[GT] {
\label{fig:b}
\includegraphics[width=0.118\columnwidth]{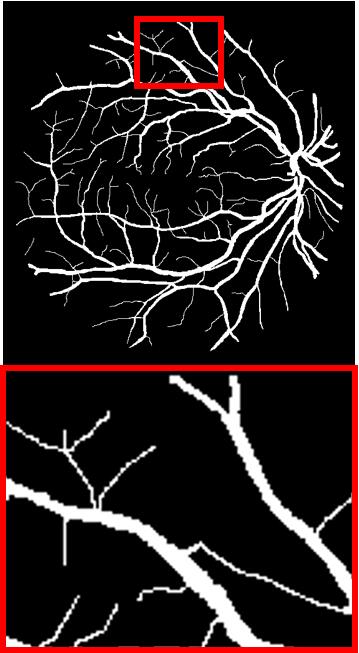}
}
\subfigure[M-Net] {
\label{fig:c}
\includegraphics[width=0.118\columnwidth]{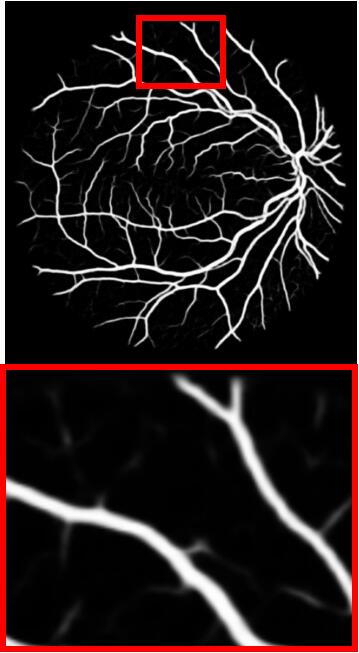}
}
\subfigure[AG-Net] {
\label{fig:c}
\includegraphics[width=0.118\columnwidth]{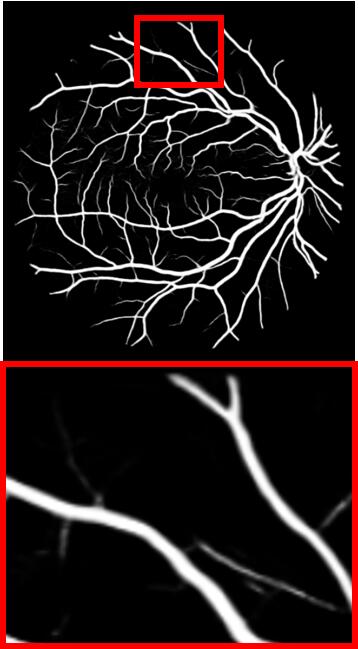}
}
\subfigure[BL] {
 \label{fig:a}
\includegraphics[width=0.118\columnwidth]{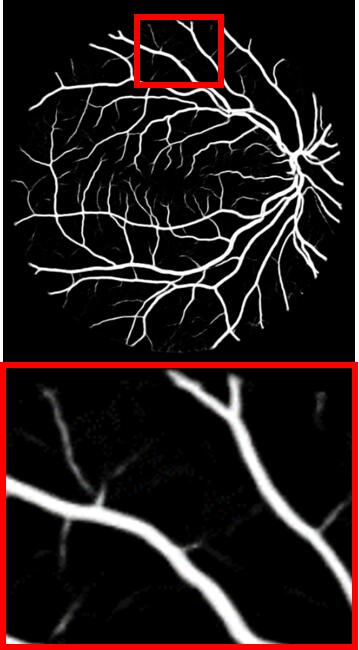}
}
\subfigure[BLST] {
 \label{fig:a}
\includegraphics[width=0.118\columnwidth]{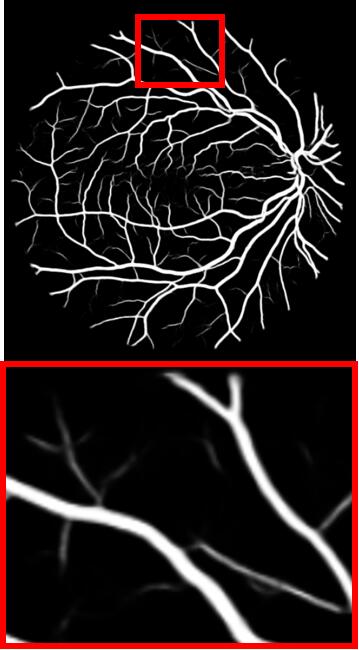}
}
\subfigure[Ours] {
 \label{fig:a}
\includegraphics[width=0.118\columnwidth]{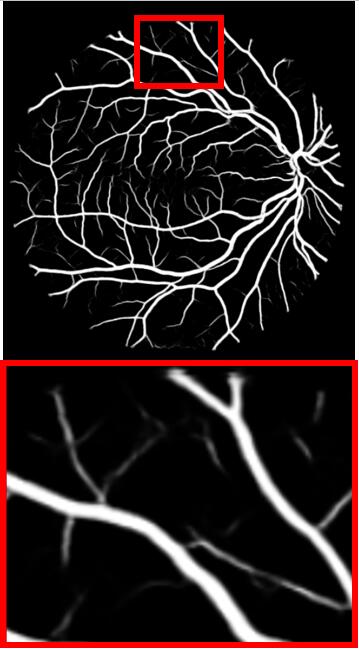}
}
\caption{ Example results for DRIVE. M-Net, AG-Net, and BL disregard some edge structures, which are very similar to textures. Conversely, by decomposing structures and textures, BLST gains better discrimination power and detects more tiny structures. Comparing (f) and (g), when adding the texture block,
%the boundaries are more clearly and
more tiny boundary structures are detected.
}
\label{figVisualCompare}
\end{figure}

We remove the texture block, structure loss $\mL_s$, and texture loss $\mL_t$ from STD-Net and name the baseline model as \textbf{BL}. The model \textbf{BLST} is formed by adding the structure-texture loss into BL. Fig. \ref{figVisualCompare} shows a test example, including the ground truth vessel (GT) and segmentation results obtained by M-Net, AG-Net, BL, BLST, and the proposed STD-Net. The experimental results of BL and BLST are shown in Table \ref{TABLE-Ablation}.

\subsection{Optic Disc/Cup Segmentation on ORIGA}
\label{Sec:ORIGAResult}
\begin{table}[!t]
	%\scriptsize
	\centering
	\caption{Comparisons of different methods with ORIGA and REFUGE}

	\begin{tabular}{l||ccc|ccc }
		\toprule
        \multicolumn{1}{c||}{ ~ }& \multicolumn{3}{c|}{ORIGA} & \multicolumn{3}{c}{REFUGE}  \\
        \midrule
		Method & $~~OE_{disc}~~$ & $~~OE_{cup}~~$ & $~~OE_{total}~~$ & $~~OE_{disc}~~$ & $~~OE_{cup}~~$ & $~~OE_{total}~~$\\
		\midrule
		ASM \cite{yin2011model} & $0.148$ & $0.313$ & $0.461$ & $-$ & $-$ & $-$\\
		SP \cite{cheng2013superpixel} & $0.102$ & $0.264$ & $0.366$  & $-$ & $-$ & $-$\\
		LRR \cite{xu2014optic} & $-$ & $0.244$ & $-$  & $-$ & $-$ & $-$\\
		U-Net \cite{ronneberger2015u} & $0.115$ & $0.287$ & $0.402$  &$0.171$ & $0.257$ & $0.428$\\
		AG-Net \cite{zhang2019attention}~~~~ & $\mathbf{0.061}$ & $0.212$ & $0.273$ &$0.178$ & $0.220$ & $0.398$ \\
		M-Net \cite{fu2018joint} & $0.071$ & $0.230$ & $0.301$  &$0.204$ & $0.231$ & $0.435$\\
		%BL        & $0.065$ & $\mathbf{0.217}$ & $\mathbf{0.282}$ \\
		%BL+ $\mL_s$ & $0.063$ & $\mathbf{0.211}$ & $\mathbf{0.274}$ \\
		%BL+ $\mL_t$ & $0.064$ & $0.213$ & $0.277$ \\
		%BLST & $0.063$ & $0.210$ & $0.273$ \\
		STD-Net & $0.063$ & $\mathbf{0.208}$ & $\mathbf{0.271}$  &$\mathbf{0.168}$ & $\mathbf{0.217}$ & $\mathbf{0.385}$ \\
		\bottomrule
	\end{tabular}
	\label{Table-ORIGA-REFUGE}
\end{table}

Optic Disc/Cup Segmentation is another important retinal segmentation task. In this experiment, we employ the ORIGA dataset, which contains 650 fundus images with 168 glaucomatous eyes and 482 normal eyes. The 650 images are divided into
325 training images and 325 testing images (including 73 glaucoma cases and 95 glaucoma cases, respectively). We crop the OD area and resize it to $256\times256$ as the input.
We compare STD-Net with several state-of-the-art methods, including ASM \cite{yin2011model}, Superpixel \cite{cheng2013superpixel}, LRR \cite{xu2014optic}, U-Net \cite{ronneberger2015u}, M-Net \cite{fu2018joint}, and AG-Net\cite{zhang2019attention}.
The ASM \cite{yin2011model} employs the circular hough transform initialization to segmentation. The superpixel method \cite{cheng2013superpixel} utilizes superpixel classification to detect the OD and OC boundaries. The method in LRR \cite{xu2014optic} obtains satisfactory results but only focuses on OC segmentation. AG-Net\cite{zhang2019attention} also strengthens the structure information but is easily influenced by the textures.

Following the setting in~\cite{zhang2019attention}, we localize the disc center with a pre-trained LinkNet \cite{chaurasia2017linknet} and then enlarge 50 pixels of bounding-boxes in up, down, right and left directions to crop the OD patch as the input image. The polar transformation is also exploited to improve the segmentation performance.
We employ overlapping error (OE) as the evaluation metric, which is defined as $OE=1-\frac{A_{GT}\bigcap A_{SR}}{A_{GT}\bigcup A_{SR}}$. $A_{GT}$ and $A_{SR}$ denote the ground truth area and segmented mask, respectively.
In particular, $OE_{disc}$ and $OE_{cup}$ are the overlapping error of OD and OE. $OE_{total}$ is the sum of $OE_{disc}$ and $OE_{cup}$.

\iffalse
\begin{table}[!t]
	%\scriptsize
	\centering
	\caption{Quantitative comparison of segmentation results on ORIGA}

	\begin{tabular}{lccc}
		\toprule
		Method & $OE_{disc}~~~$ & $OE_{cup}~~~$ & $OE_{total}$\\
		\midrule
		ASM \cite{yin2011model} & $0.148$ & $0.313$ & $0.461$\\
		SP \cite{cheng2013superpixel} & $0.102$ & $0.264$ & $0.366$\\
		LRR \cite{xu2014optic} & $-$ & $0.244$ & $-$\\
		U-Net \cite{ronneberger2015u} & $0.115$ & $0.287$ & $0.402$\\
		AG-Net \cite{zhang2019attention}~~~~ & $\mathbf{0.061}$ & $0.212$ & $0.273$ \\
		M-Net \cite{fu2018joint} & $0.071$ & $0.230$ & $0.301$\\
		STD-Net & $0.063$ & $\mathbf{0.208}$ & $\mathbf{0.271}$ \\
		\bottomrule
	\end{tabular}
	\label{Table-ORIGA}
\end{table}
\fi

Table \ref{Table-ORIGA-REFUGE} shows the segmentation results. Our method outperforms all the state-of-the-art OC segmentation algorithms, which demonstrates the effectiveness of our model. For OD segmentation, the proposed STD-Net is slightly lower than AG-Net, but STD-Net achieves the best performance on OC segmentation and better performance when considering OC and OD segmentation. Our STD-Net performs much better than the original M-Net, which further demonstrates that our structure-texture demixing method is beneficial for the segmentation performance.

We obtained similar results with the REFUGE dataset \cite{tz6e-r977-19}, which are shown in Table \ref{Table-ORIGA-REFUGE}. The training set and validation set of REFUGE have distinct appearances due to different shooting equipment, which requires a high generalization ability to reduce overfitting. Therefore, the results with REFUGE can better demonstrate the ability of structural texture decomposition.

\subsection{Ablation Study}
\label{Sec:AblationInvestigation}

We conduct an ablation investigation to further verify the effectiveness of the structure-texture demixing mechanism and texture block.
The results for DRIVE are presented in Table \ref{TABLE-Ablation}. We note several interesting observations. First, when BL considers the structure loss $\mL_s$ or the texture loss $\mL_t$, the results are improved with metrics other than Spe. With the structure loss, BL achieved the highest Sen score, which shows that more vessel structures are detected.
Second, when BL considers both the structure loss $\mL_s$ and the texture loss $\mL_t$, it achieves higher Acc, AUC and IOU scores, which demonstrates the superiority of the structure-texture demixing strategy.
Last, when BL further incorporates the texture block (STD-Net), it achieves the highest scores for Acc, AUC, and IOU. This finding demonstrates the effectiveness of the texture block. As shown in Table \ref{TABLE-Ablation},
similar results are obtained for ORIGA.

\begin{table}[!t]
	%\scriptsize
	\centering
	\caption{Ablation study with DRIVE and ORIGA}

	\begin{tabular}{l||ccccc|ccc }
		\toprule
        \multicolumn{1}{c||}{ ~ }& \multicolumn{5}{c|}{DRIVE} & \multicolumn{3}{c}{ORIGA}  \\
        \midrule
		Method & Acc~~ & AUC~~ & Sen~~ & Spe~~ & IOU~~ & $~~OE_{disc}~~$ & $~~OE_{cup}~~$ & $~~OE_{total}$ \\
		\midrule
        BL & $0.9678~~$ & $0.9829~~$ & $0.7776~~$ & $\mathbf{0.9864}~~$ & $0.6785~~$ & $0.065$ & $0.217$ & $0.282$ \\
        BL+$\mL_{s}$  & $0.9684~~$ & $0.9842~~$ & $\mathbf{0.8236}~~$ & $0.9827~~$ & $0.6948~~$ & $\mathbf{0.063}$ & $0.211$ & $0.274$ \\
        BL+$\mL_{t}$~~~   & $0.9687~~$ & $0.9841~~$ & $0.8167~~$ & $0.9837~~$ & $0.6951~~$ & $0.064$ & $0.213$ & $0.277$\\
        BLST   & $0.9691~~$ & $0.9859~~$ & $0.8201~~$ & $0.9837~~$ & $0.6984~~$ & $\mathbf{0.063}$ & $0.210$ & $0.273$\\
        STD-Net & $\mathbf{0.9695}~~$ & $\mathbf{0.9863}~~$ & $0.8151~~$ & $0.9846~~$ & $\mathbf{0.6995}~~$ & $\mathbf{0.063}$ & $\mathbf{0.208}$ & $\mathbf{0.271}$\\
		\bottomrule
	\end{tabular}
	\label{TABLE-Ablation}
\end{table}

\section{Conclusion}
\label{Sec:Conclusion}

In this paper, we have proposed a trainable structure-texture demixing network (STD-Net) to decompose an image into a structure component and texture component and separately process them. In this way, the segmentation model focuses more on structure information and reduces the influence of texture information. We have also proposed a texture block to further extract the structural information from the texture component, which substantially improves the segmentation results.
Extensive experiments for two retinal image segmentation tasks (\textit{i.e.}, blood vessel segmentation, optic disc and cup segmentation) demonstrate the effectiveness of our proposed method.

\small{
\noindent
\textbf{Acknowledments.} This work was partially supported by National Natural Science Foundation of China (NSFC) 61836003 (key project), Program for Guangdong Introducing Innovative and Enterpreneurial Teams 2017ZT07X183, Guangdong Provincial Scientific and Technological Funds under Grant 2018B010107001, Grant 2019B010155002, Tencent AI Lab Rhino-Bird Focused Research Program (No. JR201902), Fundamental Research Funds for the Central Universities D2191240.}

\bibliographystyle{IEEEtran}
\bibliography{paper1058} 

\end{document}